\documentclass[12pt,journal,onecolumn]{IEEEtran}  

\usepackage{amsmath}
\usepackage{subcaption}
\usepackage{graphicx}
\usepackage{caption}
\usepackage{subcaption}
\usepackage{tabularx,multirow}
\usepackage{enumerate}
\usepackage{makecell,textcomp}
\usepackage{url} 
\usepackage{xcolor}         
\usepackage{soul}
\usepackage{lineno}
\usepackage{textgreek}

\def\thline{\noalign{\hrule height 1.0pt}}

\begin{document}

\title{DPSNN: Spiking Neural Network for Low-Latency Streaming Speech Enhancement}

\author{\IEEEauthorblockN{Tao Sun\IEEEauthorrefmark{1},
Sander Boht\'e \IEEEauthorrefmark{2}}

\IEEEauthorblockA{Machine Learning Group,
CWI, Amsterdam, The Netherlands\\
Email: \IEEEauthorrefmark{1}tao.sun@cwi.nl,
\IEEEauthorrefmark{2}sbohte@cwi.nl}}



\maketitle

\begin{abstract}
Speech enhancement (SE) improves communication in noisy environments, affecting areas such as automatic speech recognition, hearing aids, and telecommunications. With these domains typically being power-constrained and event-based while requiring low latency, neuromorphic algorithms in the form of spiking neural networks (SNNs) have great potential. Yet, current effective SNN solutions require a contextual sampling window imposing substantial latency, typically around 32ms, too long for many applications. Inspired by Dual-Path Spiking Neural Networks (DPSNNs) in classical neural networks, we develop a two-phase time-domain streaming SNN framework -- the Dual-Path Spiking Neural Network (DPSNN). In the DPSNN, the first phase uses Spiking Convolutional Neural Networks (SCNNs) to capture global contextual information, while the second phase uses Spiking Recurrent Neural Networks (SRNNs) to focus on frequency-related features. In addition, the regularizer suppresses activation to further enhance energy efficiency of our DPSNNs. Evaluating on the VCTK and Intel DNS Datasets, we demonstrate that our approach achieves the very low latency (approximately 5ms) required for applications like hearing aids, while demonstrating excellent signal-to-noise ratio (SNR), perceptual quality, and energy efficiency.
\end{abstract}

%
%
%
%
%

\section{Introduction}
Speech enhancement refines and clarifies spoken communication in the presence of undesirable noisy conditions \cite{wang2018supervised}. Beyond automatic speech recognition (ASR) and speaker recognition, effective speech enhancement is also vital in domains such as hearing aids and mobile telecommunications. Machine learning methods, particularly deep neural networks (DNNs), in the last decade have emerged as the main approach for speech enhancement \cite{wang2013towards,xu2014regression,pandey2018confnew,gerkmann2015phase, choi2019phase,fu2017complex,luo2018tasnet,luo2019conv,luo2020dual}. 

\begin{figure*}[!ht]
     \centering
     \begin{subfigure}[b]{0.46\textwidth}
         \centering
         \includegraphics[width=\textwidth]{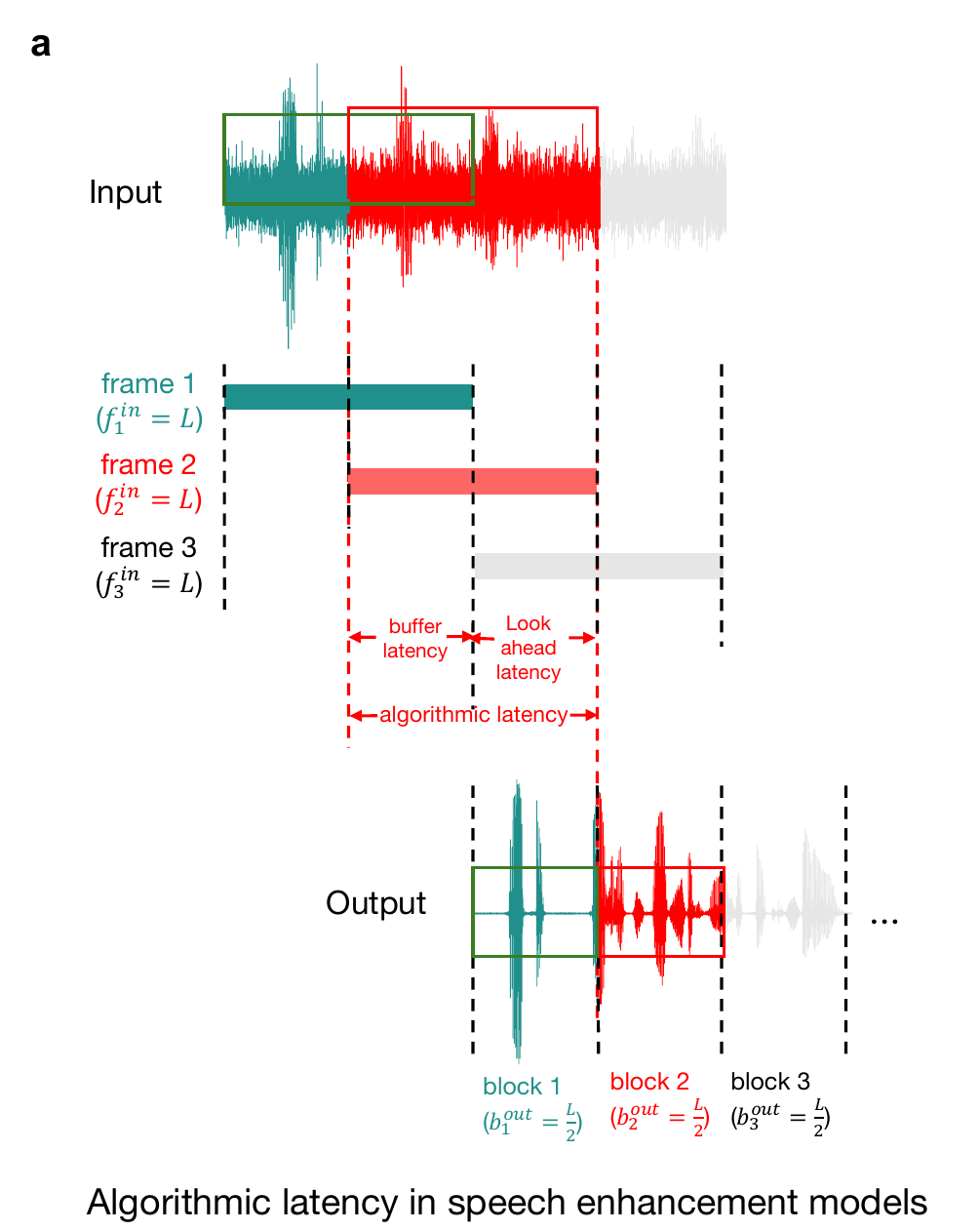}
     \end{subfigure}
     \hfill
     \begin{subfigure}[b]{0.49\textwidth}
         \centering
         \includegraphics[width=\textwidth]{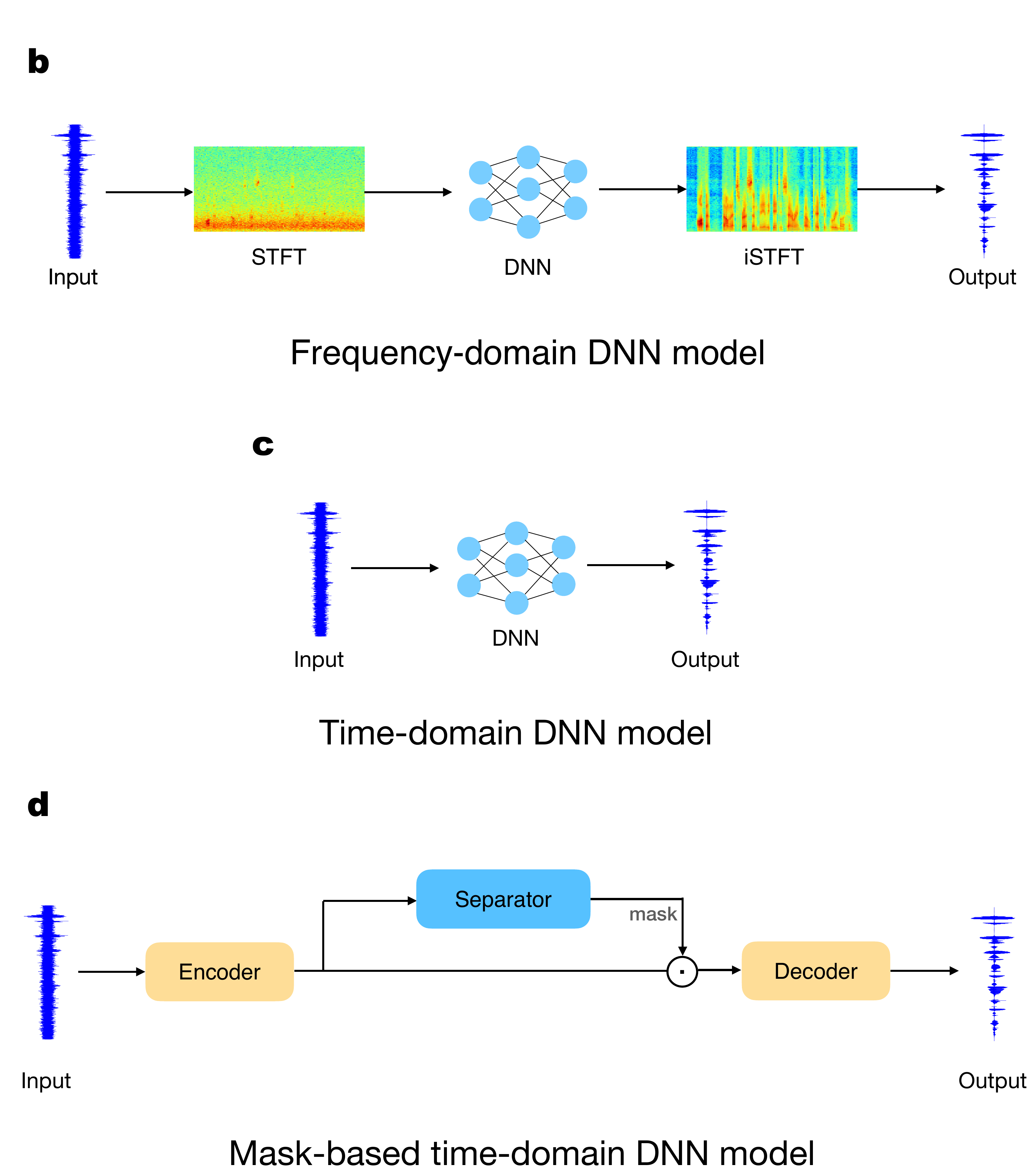}
     \end{subfigure}
     \caption{\textbf{a}, The algorithmic latency in block-based models consists of a buffering latency and a look-ahead latency. The buffering latency matches the frame-shift length (i.e. block size), whereas the look-ahead latency results from the extra look-ahead within a frame, typically used to provide additional processing context to improve performance. \textbf{b}, A frequency-domain DNN model transforms a noisy audio signal to its T-F representation by the STFT and then fed it into a neural network. \textbf{c}, Inputs and outputs to the time-domain DNN models are both time-domain signals. \textbf{d}, Mask-based time-domain DNN models commonly adopt an encoder-separator-decoder architecture.}
     \label{fig:arch}
\end{figure*}

For many SE applications, low-latency processing is critical to ensure satisfactory speech communication, like for hearing aids \cite{drgas2023survey}. Latency, also called processing latency, is defined as the delay between the input of an audio signal and the corresponding output of the processed signal (Figure \ref{fig:arch}a). Latency is decomposed into two components, algorithmic latency and hardware latency, where  algorithmic latency refers to the latency caused by algorithmic constraints, while hardware latency denotes the time needed by the hardware to process an input unit \cite{wang2022stft}. Current DNN-based SE solutions, particularly time-domain DNN models that directly handle and predict waveform signals, have achieved both high accuracy and low algorithmic latency at the same time \cite{drgas2023survey}. However, due to their large network sizes, these DNN solutions are usually energetically costly, limiting their applicability within the many power-constrained environments \cite{timcheck2023intel}.

The development of spiking neural networks (SNNs) and associated neuromorphic hardware \cite{timcheck2023intel} for speech enhancement is mainly motivated by their potential for energy efficiency, where the temporal nature of speech signal processing plays to the strengths of SNNs to handle dynamic, time-dependent tasks \cite{timcheck2023intel}. Increased energy efficiency would extend battery life and enable smaller form factors for SE devices like headsets, earbuds, hearing aids, and cochlear implants. 


Early DNN-based SE solutions typically worked in the frequency-domain, where noisy audio signals are first transformed into Time-Frequency (T-F) representations through the Short Time Fourier Transform (STFT) and then fed into a neural network (Figure \ref{fig:arch}b).
During STFT, audio signals are divided into overlapping frames, each of which is then transformed into frequency forms. A frequency-based solution typically takes an individual frame as input and produces a block as output. As shown in the Figure \ref{fig:arch}a, the length of an input frame is sum of the length of an output block and a future context. This setting implies that the production of a block is contingent upon the analysis of the entire frame. As a result, the algorithmic latency to output a block is equal to duration of a frame, which consists of duration of a block (buffering latency) and a future context (look-ahead latency) \cite{wang2022stft}. Since the frame length impacts both the time and frequency resolutions of a T-F representation, the block size and future context play an important role in determining the performance of a frequency-domain speech enhancement method. 
To trade-off between the time and frequency resolutions, and thus ensure enhancement performance, frequency-domain DNN solutions usually choose a large frame length as long as 32 ms \cite{koizumi2019trainable,wang2022stft}, which causes the SE algorithm to incur an algorithmic latency of 32 ms. Although such latency is tolerable in some applications like audio communication \cite{itu2000recommendation}, it cannot provide a satisfactory listening experience in some application scenarios, such as hearing aid, where the permissible delay tolerance ranges from 20 to 30 ms \cite{drgas2023survey}. The Clarity speech enhancement challenge \cite{claritychallenge}, targeting at hearing aid, even has the latency requirement as low as 5 ms. In contrast, time-domain models (Figure \ref{fig:arch}c), where input and output to the DNNs are both in the time-domain form, have realized low-latency enhancement by applying very short frame size; as a result, in the 2022 Clarity challenge, nearly all leading solutions were time-domain based DNN models \cite{wang2022stft}. 

 \begin{figure*}[!ht]
  \centering
     \includegraphics[width=0.95\textwidth]{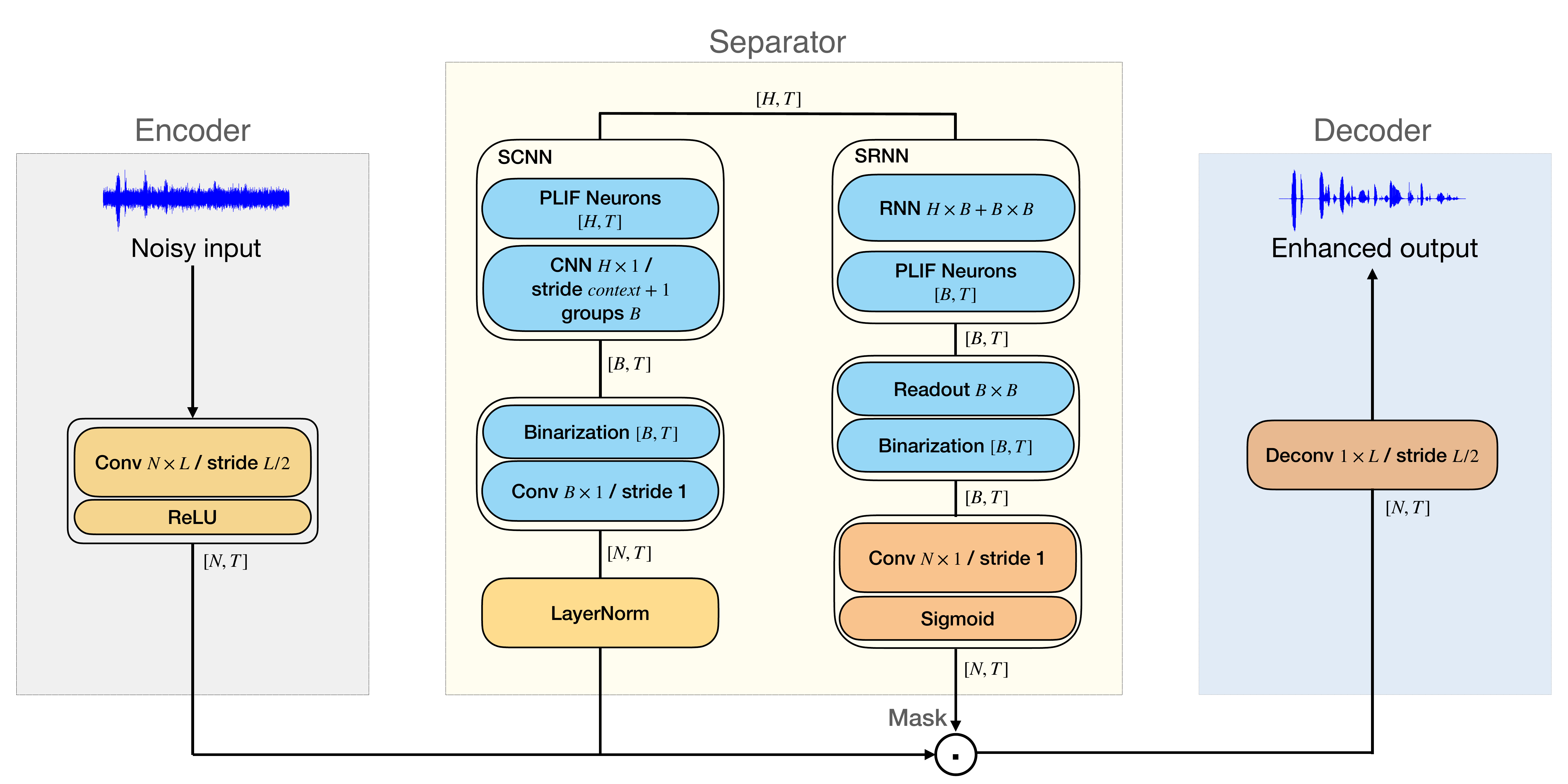}
  \caption{The proposed DPSNN adopts the encoder-separator-decoder architecture. The encoder uses convolutions to convert waveform signals into encoded 2D feature maps, effectively replacing the function of STFT. In the separator, 2D masks are calculated, primarily relying on the SCNN and SRNN modules that capture the temporal and frequency contextual information of the encoded feature maps, respectively. After applying the masks to the feature maps from the encoder, the decoder transforms the masked feature maps back to enhanced waveform signals.}

  \label{fig:arch_detail}
\end{figure*}

Yet, to the best of our knowledge, current SNN-based SE solutions are in the frequency-domain, leading to long latencies (typically 32 ms) \cite{Intel2023DNSPage,du2023spiking,riahi2023single}. While these approaches have demonstrated satisfactory enhancement performance, their prolonged latencies make them impractical for applications requiring low latency. To tackle this issue, we here design a novel time-domain streaming SNN model for speech enhancement, achieving low latency through applying small frame sizes. 
Inspired from the success of Dual-Path Recurrent Neural Networks (DPRNNs) \cite{luo2020dual}, our model applies a two-phase framework to capture rich contexts for sequence modeling.
As illustrated in Figure \ref{fig:arch_detail}, the model, dubbed Dual-Path Spiking Neural Network (DPSNN), first converts noisy signals into a 2D feature map via a convolution that functions similarly to the STFT, but with a significantly smaller kernel size, leading to much lower latency. A DPSNN then consists of two distinct modules, a Spiking Convolutional Neural Network (SCNN) module and a module based on Spiking Recurrent Neural Networks (SRNN) \cite{yin2021accurate}, to separate clean signals from noisy inputs. The initial phase of a DPSNN works with the SCNN module and captures contextual information along the temporal direction of the 2D feature map. The subsequent phase, via  the SRNN module, then integrates context along the frequency direction of the 2D feature map. Taken together, this approach allows the use of a small frame size and achieves correspondingly low latency. Additionally, our DPSNN applies threshold-based and regularizer-based activation suppressions \cite{zhu2023star} to specific non-spiking layers, creating more sparse representations and thus enhancing energy-efficiency of the DPSNN model. 

We conduct thorough evaluations of proposed DPSNNs using experiments on the VCTK database \cite{valentini2017noisy} and Intel DNS Dataset \cite{timcheck2023intel}, yielding excellent performance across latency, energy efficiency, and speech enhancement metrics.

\section{Related work}
\subsection{Time-domain solutions}
Time-domain approaches can be divided into mapping-based and mask-based methods \cite{sun2022time}. Mapping-based models typically apply an encoder-decoder architecture, directly mapping noisy inputs to denoised outputs  \cite{pandey2019tcnn,tsao_2017_c,gong2019dilated,macartney2018improved}. Mask-based methods (Figure \ref{fig:arch}d), and in particular the TasNets (Time-domain Audio Separation Network) models \cite{luo2018tasnet,luo2019conv,luo2020dual}, adopt an encoder-separator-decoder architecture and have demonstrated excellent performance in both separation and latency metrics \cite{drgas2023survey,wang2022stft}. In their encoders, a trained convolutional layer 
functions similar to an STFT to generate 2D spatiotemporal features, from which masks are computed in the separator. These masks are then multiplied with the spatiotemporal features to produce enhanced features. Lastly, the separated enhanced features are reverted to waveforms in the decoder which is comprised of a deconvolution layer. In the mask-based method, the kernel size 
of the convolution in the encoder is equivalent functionally to the frame size 
of the STFT. As such, the latency of a TasNet is determined by the kernel size used in its encoder. In the original TasNet framework \cite{luo2018tasnet}, LSTM networks were employed within the separator to compute the mask. To alleviate the computational burden associated with such LSTM networks, Conv-TasNet \cite{luo2019conv} 
proposed the use of dilated convolutions in place of LSTMs within the separator, allowing for smaller kernel sizes and strides in convolutions within the encoder, enhancing its suitability for low-latency applications. To also effectively capture long sequential inputs, DPRNNs (Dual-Path RNNs) were introduced \cite{luo2020dual}, 
the separator of which consisted of two sequential RNNs processing shorter chunks of input signals: the first RNN operates within each chunk in parallel, integrating local frequency-related contexts, while the second RNN operates across chunks to capture long-term temporal information.

\subsection{Speech enhancement with SNNs}
The Intel Neuromorphic Deep Noise Suppression Challenge (Intel N-DNS Challenge) focused on speech enhancement tasks as a high potential application domain. The Challenge winner, Spiking-FullSubNet, combined two frequency domain approaches, using a full-band model and a sub-band model \cite{Clairaudience2024speech}. The full-band model here captured dependencies between frequency bands, while the sub-band model handled each band independently. Additionally, the Gated Spiking Neuron (GSN) was introduced in this model, where membrane potentials are calculated through time constants that could vary at each time step. Another competitive approach, Spiking Structured State Space Model (SpikingS4) \cite{du2023spiking}, builds on the concept of structured state space modeling, while the Spiking-UNet \cite{riahi2023single} combines the UNet architecture with SNNs for single-channel noise reduction. 
Despite their strong performance in speech enhancement metrics, these SNN models, being frequency-domain methods, are unable to meet low-latency requirements in demanding application scenarios (e.g., hearing aid) without substantial modifications. \cite{wang2022stft}.

\section{Methods}
\subsection{Problem setup}
Speech enhancement improves the quality of speech signals by reducing or eliminating additive noise. The primary goal is to enhance the intelligibility and perceptual quality of speech in various real-world environments where noise interference is present. 

One common approach to formally modeling speech enhancement involves the use of signal processing techniques to model the relationship between the observed noisy speech signal $y[n]$, the clean speech signal $s[n]$, and the additive noise signal $v[n]$. The noisy signal $y[n]$ can be expressed as the sum of the clean speech signal and the additive noise:
\begin{gather}
y[n]=s[n]+v[n].
\end{gather}
The goal of speech enhancement is then to estimate or reconstruct the clean speech signal 
$s[n]$ from the observed noisy signal $y[n]$. The enhanced speech signal $\tilde{s}[n]$ is written as:
\begin{gather}
\tilde{s}[n]=f(y[n]).
\end{gather}
Speech enhancement is closely related to speech separation;  the key difference being that speech enhancement aims to improve the quality of a noisy speech signal by removing or suppressing the unwanted noise, whereas speech separation focuses on separating individual speakers or sound sources from a mixture of multiple audio sources.

\subsection{Spiking neural networks (SNNs)}
SNNs typically use similar network topologies as ANNs, yet SNNs employ stateful, binary-valued spiking neurons as their computational units. Consequently, inference in SNNs unfolds iteratively across multiple time steps $t = 0, 1, ..., T$: at each time step $t$, the internal state in the form of the membrane potential $u_t$ is influenced by incoming spikes from connected neurons emitted at time step $t-1$, as well as the neuron's previous potential $u_{t-1}$. Upon reaching a threshold $\theta$, the neuron emits a spike, and typically does so sparingly. This sparse and asynchronous communication among connected neurons  enables SNNs to potentially achieve high energy-efficiency.

Various spiking neuron models exist, ranging from the intricate and biologically detailed Hodgkin-Huxley model to the simplified Leaky-Integrate-and-Fire (LIF) neuron model \cite{Gerstner2002-wd}. For machine learning applications, SNNs mostly employ LIF spiking neurons, and variants thereof, due to their interpretability and computational efficiency. Resembling an RC circuit, the LIF neural model is mathematically represented as:
\begin{equation} \label{eq:lif_diff}
\tau_{m}\frac{du}{dt} = -(u-u_{rest})  + RI,
\end{equation}
where $u_{rest}$ denotes the resting potential of the neuron, $I$ expresses the input current, R is the membrane resistance, and $\tau_m$ represents the membrane time constant. 


The discrete approximation of (\ref{eq:lif_diff}) can be expressed as
\begin{alignat}{3}
u_t &= \left(1-\frac{1}{\tau_m}\right)u_{t-1} + \frac{1}{\tau_m} \left(u_{rest} + RI_t\right) \label{eq:subthreshold}\\
s_t &= \Theta\left(u_t-\theta\right)  \label{eq:spike}\\
u_t &= u_t(1-s_t) + u_{rest}s_t \label{eq:thresholded}
\end{alignat}
where equation (\ref{eq:subthreshold}) describes subthreshold neural dynamics of a neuron; $I_t=\Sigma_{i}w_{i}s_{t-1}^i$ is the input current from the pre-synaptic neurons, where $s_{t-1}^i$ indicates whether a pre-synaptic neuron $i$ spikes in the last time step $t-1$; $\Theta$ in (\ref{eq:spike}) is Heaviside step function deciding whether a neuron spikes; equation (\ref{eq:thresholded}) calculates the final potential of a neuron in a time step.

For training SNNs, Surrogate Gradient (SG) methods \cite{neftci2019surrogate,yin2021accurate} and LIF neurons with learnable model parameters \cite{yin2021accurate,fang2021incorporating} have enhanced the performance of SNNs and enabled straightforward supervised trainability. In \cite{fang2021incorporating}, Parametric LIF (PLIF) neurons are introduced where the time constant $\tau_m$ of a LIF neuron is learnable and shared by all neurons in one layer. In PLIF neurons, to ensure $1-\frac{1}{\tau_m} < 1$, $\tau_m$ is calculated through a sigmoid function, 
\begin{equation} 
\tau_m = \frac{1}{k(a)} = \frac{1}{1+e^{-a}}.
\end{equation}

In \cite{yin2020effective}, adaptive LIF (ALIF) neurons used, where both time constants $\tau_m$ and $\tau_adp$ are learnable in each individual LIF neuron. Additionally, the threshold of a neuron increases after spiking and then decays with a time constant $\tau_{adp}$. Dynamics in ALIF neurons can be written as
\begin{align}
\label{eq2:adapt}
&\eta_t = \rho \eta_{t-1}+ (1-\rho)s_{t-1} \\
&\theta = b_0+\beta\eta_t  \label{eq:dynamic_threshold}\\
&u_t = \alpha u_{t-1} + (1-\alpha)R I_t - s_{t-1}\theta ,
\label{eq2:adaptsn}
\end{align}
where $\alpha = \exp(-dt/\tau_m)$ and $\rho = \exp(-dt/\tau_{adp})$ express decay of the membrane potential and threshold, respectively. In equation \eqref{eq:dynamic_threshold}, $\theta$ is a dynamical threshold consisting of a constant minimal $b_0$ and an adaptive part $\eta_t$, whose coefficient $\beta$ is a constant controlling adaptation of the threshold.

\subsection{Architecture}

\begin{figure*}[t!]
  \centering
  \includegraphics[width=0.8\textwidth]{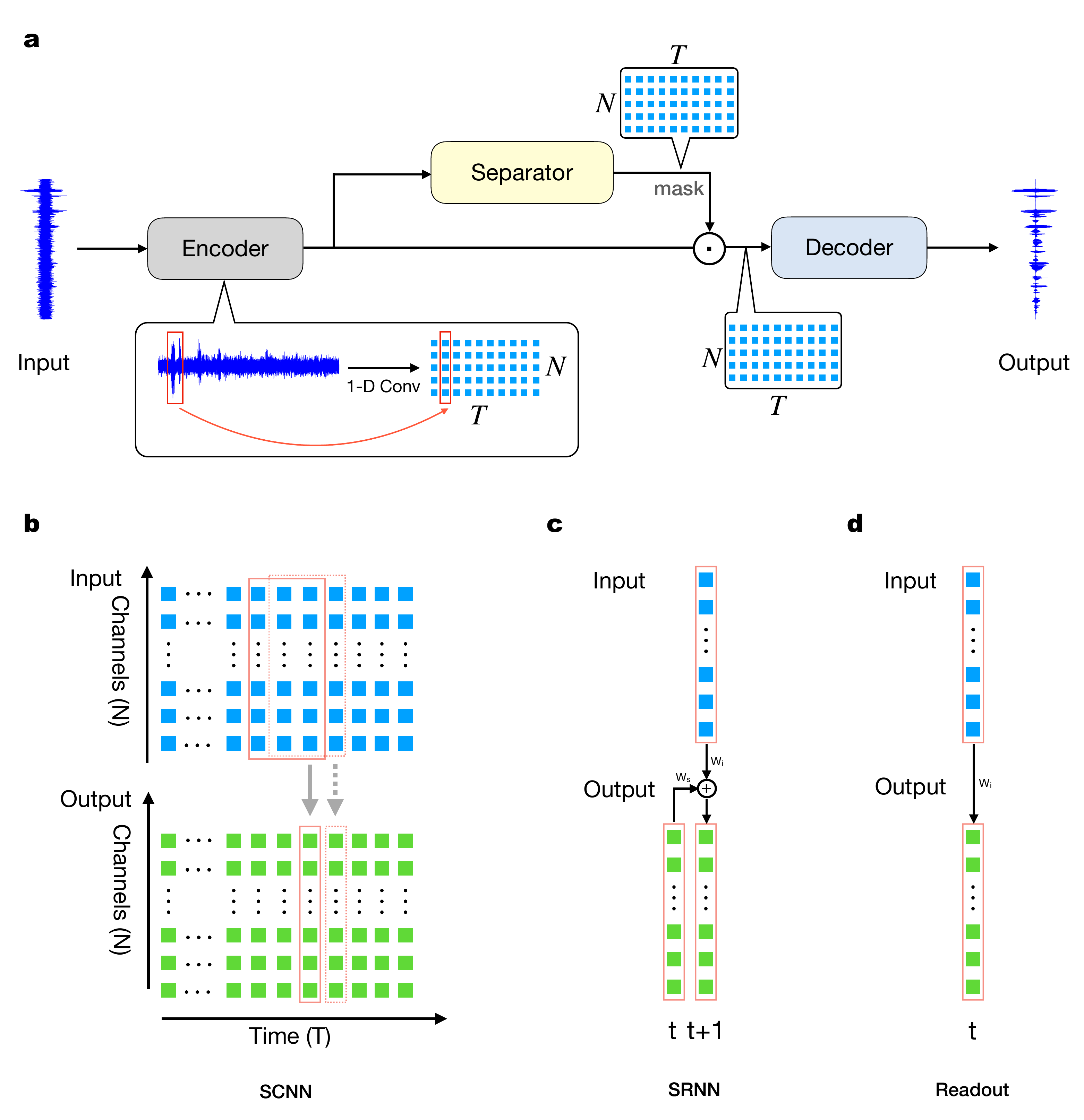}
  \caption{\textbf{a}, In the mask-based encoder-separator-decoder  architecture, the encoder converts overlapping frames into 1D features through convolution and aligns them into a 2D feature map. Each 1D feature is processed in one time step in the subsequent spiking layers. \textbf{b}, In the SCNN layer, a group convolution is applied along the temporal axis of a feature map to capture temporal contextual information. \textbf{c}, The SRNN layer is a fully-connected recurrent spiking layer that integrates contexts along the frequency direction of its input 2D feature map. \textbf{d}, Readout is done using a fully-connected readout layer with non-spiking neurons, where the membrane potential of these neurons is calculated and output without any spiking or resetting.}
  \label{fig:arch-blocks}
\end{figure*}

As shown in Figure \ref{fig:arch-blocks}a, our speech enhancement model adopts the mask-based encoder-separator-decoder architecture. An encoder first takes overlapping waveform frames, each with a length of $L$, as inputs and converts these frames into features with $N$ channels through a linear transformation, specifically a one-dimensional (1D) convolution and a ReLU activation function. Those features,  aligned together, then form a two-dimensional (2D) feature map. 


The separator learns a mask to extract clean signals from noisy inputs. This is achieved using two spiking modules, SCNN and SRNN (Figure \ref{fig:arch-blocks}b and \ref{fig:arch-blocks}c, respectively). Initially, the 2D feature map generated by the encoder passes through a layer normalization layer, followed by a bottleneck $1\times 1$ convolution layer, resulting in an output with $B$ channels. To reduce computations in the subsequent SCNN layer, an activation suppression operation \cite{zhu2023star} with a learnable threshold is applied to binarize the output, converting values below the threshold to zeros and values above it to ones. After processing through the SCNN and SRNN layers, the output is fed into a fully-connected readout layer using non-spiking neurons, as illustrated in Figure \ref{fig:arch-blocks}d. The readout layer's activations are also subjected to suppression by a learnable threshold, converting values below it to zeros while keeping values above it intact. This suppressed output is then passed through a $1\times 1$ convolution layer with a Sigmoid activation function to produce a mask for clean signals. Finally, the input features, multiplied by the mask computed by the separator, are converted back to a one-dimensional (1D) enhanced signal by the encoder, which comprises of a deconvolutional layer.

The following provides more details on the SCNN and SRNN layers.


\paragraph{SCNN layer}
The SCNN layer takes in the features output by the binarized bottleneck layer and integrates context along its temporal direction. Specifically, a group convolution is carried out along the temporal axis of a feature map ($b$ channels) to integrate contexts of a predefined number of previous time steps (Figure \ref{fig:arch-blocks}b), producing an output feature map with $H$ channels. For one time step of this layer, each input channel is convolved with its own set of $\frac{H}{B}$ filters; the input feature map is zero-padded before the first time step. Within this module, we use PLIF neurons \cite{fang2021incorporating}. To train PLIF neurons, the surrogate gradient function introduced in \cite{fang2021incorporating} is used, where we use as surrogate gradient the function $\sigma(x) = \frac{1}{\pi} \arctan(\pi x) + \frac{1}{2}$.

\paragraph{SRNN layer}
The SRNN layer, based on the SRNNs introduced in \cite{yin2021accurate}, is a fully-connected recurrent layer of spiking neurons that captures contexts along the frequency direction to extract frequency-related features. Taking the output from the SCNN layer, it produces outputs with $B$ channels. We use adaptive LIF (ALIF) neurons \cite{yin2020effective} in this layer; for training, the multi-Gaussian surrogate gradient \cite{yin2021accurate} is used.  




In this design, the SCNN layer captures contextual information across the temporal axis of the encoder's feature map, while the SRNN layer integrates context along the frequency axis. As we will demonstrate, together they form a robust separator capable of precise mask generation, particularly effective for handling extended audio sequences while incurring low-latency.


\section{Experiments}
We assess the robustness of a speech processing model through training and testing on two primary datasets: the VCTK dataset and the Intel DNS dataset based on the MS DNS challenge dataset. The experiments encompassed both monolingual and multilingual aspects and considered various noise conditions.

\subsection{Datasets}
\paragraph{VCTK Datasets}
The VCTK Database encompasses 10 hours of speech data, with most utterances lasting no more than 5 seconds, and some as brief as 2-3 seconds. The training set, downsampled from 48kHz to 16kHz, comprises 11,575 sentences. This training dataset involves 28 speakers (14 male and 14 female), all sharing the same English accent region. Each speaker contributes around 400 sentences, providing a diverse and representative collection. The training dataset includes a set of 10 noises consisting of 2 artificially generated noise signals (speech-shaped noise and babble) and 8 real noises sourced from the Demand database. Four Signal-to-Noise Ratios (SNRs) are considered: 15 dB, 10 dB, 5 dB, and 0 dB, resulting in 40 distinct noise conditions for comprehensive training. The testing set for VCTK consists of 827 sentences and includes two speakers (one male and one female). To simulate real-world conditions, five additional noises from the Demand database are introduced under four SNRs: 17.5 dB, 12.5 dB, 7.5 dB, and 2.5 dB, yielding a total of 20 noise conditions for evaluation. For VCTK, we compare to the Spiking-UNet \cite{riahi2023single}, the sole SNN model reporting results on this dataset.

\paragraph{Intel DNS Dataset}
The Intel DNS Dataset, derived from the MS DNS Challenge dataset \cite{timcheck2023intel}, contains 500 hours of speech data distributed across both training and validation sets. Each set comprises of 60,000 samples. The dataset incorporates a range of Signal-to-Noise Ratios (SNRs) from 20 dB to -5 dB, providing varied and challenging scenarios. Each utterance within the Intel DNS dataset maintains a fixed duration of 30 seconds. Speech samples include English, German, French, Spanish, and Russian. 

\subsection{Evaluation metrics}
To assess the performance of our speech processing model, we use two distinct types of evaluation metrics: Signal-to-Noise Ratio (SNR) metrics and Perception metrics.

\subsubsection{SNR Metrics}
For the evaluation of Signal-to-Noise Ratios, we primarily use the Scale-invariant Signal-to-Noise Ratio (SI-SNR) \cite{le2019sdr}. Crucially, SI-SNR exhibits scale invariance, meaning alterations to the overall magnitude (volume) of the output do not impact SI-SNR. 

The SI-SNR is defined as:
\begin{equation}  
        \text{SI-SNR} := 10log_{10}\frac{||\mathbf{s}_{target}||_{2}}{||\mathbf{e}_{noise}||_{2}}, 
\end{equation}
where $\mathbf{s}_{target} := \frac{\mathbf{\tilde{s}} \cdot \mathbf{s}}{||\mathbf{s}||_{2}^2} \mathbf{s}$  and $\mathbf{e}_{noise} := \mathbf{\tilde{s}} - \mathbf{s}_{target}$.

\subsubsection{Perceptual Metrics}
The first perceptual metric we use is the Perceptual Estimation of Speech
Quality (PESQ) \cite{rix2001perceptual}. PESQ evaluates speech quality by comparing the original and processed speech signals at the perceptual level, taking into account factors such as distortion, noise, and speech intelligibility. It ranges from $-0.5$ to $4.5$, with higher scores indicating better perceived quality.

We also calculate DNSMOS (Distributed Network Speech Mean Opinion Score) to evaluate perceptual quality. Mean Opinion Score (MOS) serves as a perceptual metric that utilizes human evaluations to generate an average opinion score to evaluate perceived speech quality. MOS scores range from 1 to 5, where 1 signifies poor quality and 5 denotes excellent quality. In DNSMOS, a deep network is trained to predict the perceptual quality score, aligning the human perceptual quality expressed in MOS within its training corpus.

\subsubsection{Intelligibility Metric}
The Short-Time Objective Intelligibility (STOI) metric \cite{taal2011algorithm} is a method used to assess the intelligibility of speech signals. STOI is computed by dividing the short-time cross-correlation between the clean and degraded signals by the geometric mean of their short-time energies. The resulting value ranges from 0 to 1, with 1 indicating perfect intelligibility and 0 indicating no intelligibility.

\subsubsection{Power Metrics}
We apply two power metrics, namely the power proxy and power delay product (PDP) proxy, both introduced in the Intel N-DNS challenge \cite{Intel2023DNSPage}. The power proxy evaluates the effective number of synaptic operations per second, while the PDP proxy combines both latency and power efficiency into a single value. Among these metrics, the PDP proxy enables comparisons among solutions that manage trade-offs between latency and power consumption. 
For frequency-domain DNN models, only the power consumption of the DNN itself is calculated, excluding the power consumption of the STFT and iSTFT input-output conversions. When comparing with these models, we also exclude the power consumption of the encoder and decoder, the counterparts of STFT and iSTFT, in our DPSNNs.

\subsection{Training}
The loss function used to training our model consists of three components. The first component ($L_{si\text{-}snr}$) maximizes SI-SNRs of output waveforms. The second component ($L_{mse}$) is a Mean-Square-Error loss that minimizes the squared $L2$ norm between output waveforms and clean waveforms. The third component consists of the $L_1$ regularizers \cite{zhu2023star} ($ L_{1_bn}$ and $ L_{1_ro}$) that penalize the non-zeros activations in the suppressed outputs of the bottleneck convolution layer and the readout layer, respectively. Overall, the loss function is:
\begin{equation}  
 L = 100 + L_{si\text{-}snr} + 0.001 * L_{mse} + \lambda_2 * L_{1_bn} + \lambda_2 * L_{1_ro},
\end{equation}
where $\lambda_2$ and $\lambda_3$ are determined by grid-search. In our experiments, we found $0.001$ to be a good value for both.

\subsection{Optimizing the network parameters on VCTK}
\label{sec:param}
We assess the performance of our model on the VCTK dataset with channel combinations $H$,$B$ and $N$ and present the results in Table \ref{tab:config}. From our evaluation, we draw the following conclusions:
\begin{enumerate}[(i)]
    \item Encoder/decoder: Expanding the number of channels enhances spectral resolution, thereby improving overall performance.
    \item Channels in the separator: Employing a small bottleneck size $B$ alongside a large number of channels $H$ within the spiking block(s) proves effective. Additionally, larger $B$ values consistently outperform smaller ones, a deviation from the findings in \cite{luo2018tasnet} where the optimal $H/B$ value was found to be around 5. This discrepancy may be attributed to the nature of SNNs, which produce binary outputs and therefore require more neurons to convey information.
\end{enumerate}

\begin{table*}[!htbp]
	\small
	\centering
	\caption{The effect of different configurations. The length of input examples are 1 second. The filter length in the encoder $L=80$, resulting in a 5ms latency. Number of the context step is 4.}
	\vspace{0.2cm}
	\label{tab:config}
	\begin{tabular}{c|c|c|c|c|c|c|c|c|c}
		\thline
            \textbf{} & \textbf{} & \textbf{} & \textbf{} & \textbf{} & \multicolumn{3}{c|}{DNSMOS} & \textbf{} & \textbf{} \\
            \cline{6-8} 
		\thead{$N$} & \thead{$B$} & \thead{$H$} & \thead{SI-SNR} & \thead{PESQ} & \thead{\textit{OVRL}} & \thead{\textit{SIG}} & \thead{\textit{BAK}} & \thead{Params} & \thead{Learning \\ rate}\\ 
		\hline
		256 & 256 & 256 & 18.29  & 2.30 & 2.81 & 3.22 & 3.62  & 372K & 1e-2  \\
            \hline
            
            512 & 128 & 512 & 18.26 & 2.30 & 2.79 & 3.21 & 3.61 & 317K & 1e-2   \\
            
            512 & 256 & 512 & 18.34 & 2.32 & 2.85 & 3.23 & 3.73 & 613K & 7.5e-3   \\
           
            512 & 512 & 512 & 18.48 & 2.36 & 2.86 & 3.24 & 3.72 & 1.4M & 7.5e-3   \\
            
		\thline
	\end{tabular}
\end{table*}

\begin{figure}[t!]
\centering
\includegraphics[width=0.75\textwidth]{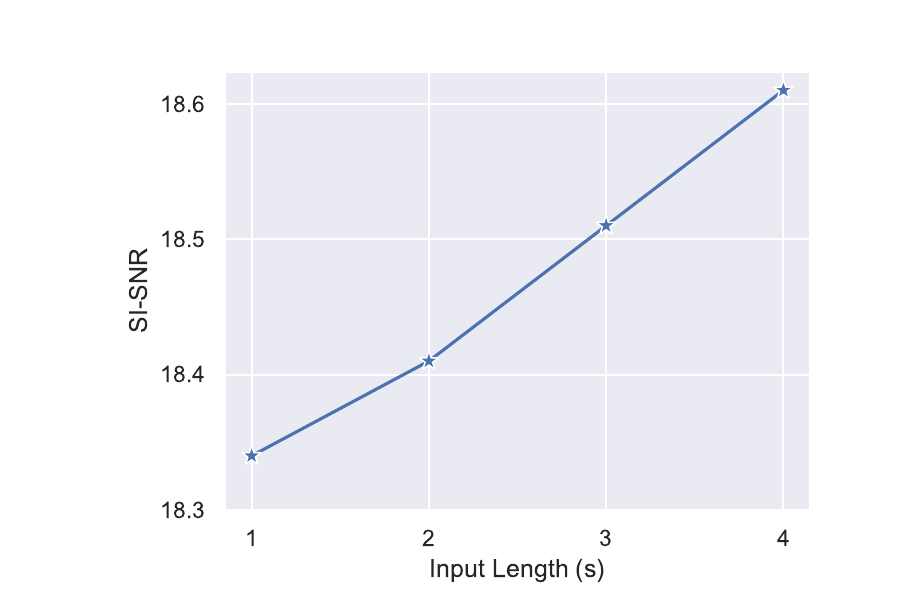}
\caption{Influence of the length of input examples on model performance. The channels in the model are $N=512, B=256, H=512$. The filter length in the encoder is $L=80$, resulting in a 5ms latency. The size of the context step is 4.} 
\label{fig:seqlen}
\end{figure}

Examining our approach in detail, we find that the performance of our model is significantly influenced by the length of input examples, both for training and for evaluation. We explored our DPSNN models with input lengths from 1.0 second to 4.0 seconds on VCTK. As illustrated in Figure \ref{fig:seqlen}, the model performs better with longer input lengths, achieving the best SI-SNR result with the 4.0-second example length.

We further assessed the impact of the context steps in the SCNN module on performance. As shown in Table \ref{tab:contextdur}
, models with four context steps generally achieve better results than those with two or eight context steps, both in terms of SNR metrics and perceptual metrics.

\begin{table}[!htbp]
	\small
	\centering
	\caption{Influence of context steps in SCNN on the model performance. The length of input examples are 1 second. The channels in the model are $N=512, B=256, H=512$.}
	\vspace{0.2cm}
	\label{tab:result-system}
	\begin{tabular}{|c|c|c|c|c|c|}
            \hline
            \textbf{} & \textbf{} & \textbf{} & \multicolumn{3}{c|}{DNSMOS}  \\
            \cline{4-6}
		\thead{Context steps } & \thead{SI-SNR} & \thead{PESQ} &\thead{\textit{OVRL}} & \thead{\textit{SIG}} & \thead{\textit{BAK}}  \\
		\hline
            2  & 18.28 & 2.29 & 2.81 & 3.20 & 3.67            \\
            4 & \textbf{18.34} & \textbf{2.32} & \textbf{2.85} & \textbf{3.23} & \textbf{3.73} \\
            8 & 18.16  & 2.29 & 2.79  & 3.21 & 3.60  \\

		\thline
	\end{tabular}
 \label{tab:contextdur}
\end{table}

Finally, we conducted experiments to assess models with varying filter lengths in the encoder ($L$). As can be seen in Table \ref{tab:winsize}, models with larger $L$ exhibit improved SI-SNRs. We attribute this enhancement to the larger context provided by a larger $L$ for each time step within the spiking block(s). However, a larger filter length also  leads to higher model latency. Moreover, it seems that filter lengths do not significantly influence performance in terms of PESQ and DNSMOS. Hence, to strike a balance between speech performance and latency, we selected $L=80$ for most of our experiments.

\begin{table}[!htbp]
	\small
	\centering
	\caption{Influences of filter lengths in the encoder ($L$) on model performance. The length of input examples are 1 second. The channels in the model are $N=512, B=256, H=512$. Number of the context step is 4.}
	\vspace{0.2cm}
	\label{tab:result-system}
	\begin{tabular}{|c|c|c|c|c|c|c|}
            \hline
            \textbf{} & \textbf{} & \textbf{} & \textbf{} & \multicolumn{3}{c|}{DNSMOS}  \\
            \cline{5-7}
		\thead{L} & \thead{Latency  \\ (ms)} & \thead{SI-SNR} & \thead{PESQ} &\thead{\textit{OVRL}} & \thead{\textit{SIG}} & \thead{\textit{BAK}}  \\
		\hline
            40  & 2.5& 17.65  & 2.24 & 2.80 & 3.20 &  3.66           \\
            80 & 5  & 18.34 & 2.32 & 2.85 & 3.23 & 3.73  \\
            160 & 10 & 18.38 & 2.28 & 2.79 & 3.20 & 3.62  \\

		\thline
	\end{tabular}
 \label{tab:winsize}
\end{table}

\subsection{Comparison of DPSNN with previous methods}
The comparison of the best performing DPSNN model with previous SNN models on VCTK is presented in Table \ref{tab:vctk}. First, we find that our model incurs significantly lower latency (5ms) compared to both the SDNN baseline model \cite{riahi2023single} and Spiking-UNet \cite{riahi2023single}. Furthermore, our model outperforms in terms of both DNSMOS and STOI metrics, while the PESQ performance falls slightly short compared to Spiking-UNet. This disparity may be attributed to the reliance of PESQ computation on the magnitude spectrogram of speech, as discussed in \cite{luo2019conv}, where a similar observation was made.

\begin{table*}[htbp]
\caption{Evaluation metrics comparisons of the VCTK dataset. The length of input examples are 4 seconds. The channels in the model are $N=512, B=256, H=512$.}
\label{tab:results}
\begin{center}
    \begin{tabular}{|c|c|c|c|c|c|c|}
    \hline
    \textbf{} & \textbf{} & \textbf{} & \multicolumn{3}{c|}{\textbf{DNSMOS}} & \textbf{} \\
    \cline{4-6} 
    \textbf{Model} & \thead{\textbf{Latency}\\ (ms)}  & \textbf{PESQ} & \textbf{\textit{OVRL}} & \textbf{\textit{SIG}} & \textbf{\textit{BAK}} & \textbf{STOI}  \\
    \hline
    Noisy                             & --   & 1.97 &2.69 & 3.34 & 3.12 & -- \\
    \hline
    SDNN baseline \cite{riahi2023single}  & 32   & 2.00 &      2.44 & 3.05 & 3.09 & 0.91\\
    Spiking-UNet \cite{riahi2023single}  &  32  & \textbf{2.66} &2.81 & 3.13 & \textbf{3.85} & 0.92 \\
     \hline
    Ours & \textbf{5}  & 2.37  & \textbf{2.94} & \textbf{3.27} & 3.84 & \textbf{0.93} \\
    \hline
    \multicolumn{5}{l}{}
    \end{tabular}
\label{tab:vctk}
\end{center}
\end{table*}

\subsection{Ablation study}
We furthermore conducted experiments on VCTK to evaluate the impact of the SCNN and SRNN layers on our model's performance. Specifically, we remove either the SCNN or SRNN layer -- the results are shown in Table \ref{tab:ablation}. We find substantial decline in overall performance in either case, in particular when removing the SCNN module. 

\begin{table}[!htbp]
	\small
	\centering
	\caption{Ablation study. The length of input examples is 1 second. The channels in the model are $N=512, B=512, H=512$. The filter length in the encoder $L=80$, resulting in a 5ms latency. Number of the context step is 4.}
	\vspace{0.2cm}
	\label{tab:result-system}
	\begin{tabular}{|c|c|c|c|c|c|}
            \hline
            \textbf{} & \textbf{} & \textbf{} & \multicolumn{3}{c|}{DNSMOS}  \\
            \cline{4-6}
		\thead{Ablation} & \thead{SI-SNR} & \thead{PESQ} &\thead{\textit{OVRL}} & \thead{\textit{SIG}} & \thead{\textit{BAK}}  \\
		\hline
		  DPSNN  & 18.48 & 2.36 & 2.86 & 3.24 & 3.72   \\
            w/o SCNN  & 17.72 & 2.18 & 2.68 & 3.19 & 3.37          \\
            w/o SRNN  & 18.38 & 2.32 & 2.82 & 3.22 & 3.67  \\

		\thline
	\end{tabular}
 \label{tab:ablation}
\end{table}

\subsection{Intel DNS Dataset}
For the Intel N-DNS challenge dataset, we compare the DPSNN  with several benchmarks, including the SDNN baseline \cite{timcheck2023intel}, the winning models of track 1 (Spiking-FullSubNet (Large) and Spiking-FullSubNet (Small) \cite{hao2024audio}), the runner-up model (CTDNN), and the Spiking S4 model \cite{du2023spiking}. Regarding speech metrics, while our best SI-SNR and SI-SNRi surpass those of Spiking S4, Spiking-FullSubNet (Small), and CTDNN, our DNSMOS performances outperforms that of Spiking S4. Most notably, our 5ms latency significantly outperforms all these models, which exhibit a latency of 32ms. The power metrics (power proxy and PDP proxy) and parameter size for the DPSNN exclude the power consumption of the encoder and decoder -- the inclusive numbers are noted in the parenthesis. 

By jointly evaluating latency and power efficiency using the PDP proxy, our DPSNN models significantly outperforms both leading Spiking-FullSubNet models. While the SDNN baseline and CTDNN have lower values in this specific metric, their speech metrics (SI-SNR and/or DNSMOS) fall well short of those achieved by our DPSNN models. This highlights the combined capability of DPSNNs in terms of speech quality, latency, and power consumption.

\begin{table*}[htbp]
\caption{Evaluation metrics comparisons on the Intel DNS dataset. The filter length (L) in the encoder of DPSNNs is 80. The channels in the models are $N=512, B=512, H=512$. The context step is 12.}
\label{tab:results}
\begin{center}
    \begin{tabular}{|c|c|c|c|c|c|c|c|c|c|}
    \hline
    \textbf{} & \textbf{} & \textbf{} & \textbf{} & \multicolumn{3}{c|}{\textbf{DNSMOS}} & \textbf{} & \textbf{} & \textbf{} \\
    \cline{5-7} 
    \textbf{Model} & \thead{\textbf{Latency}\\ (ms)} & \textbf{SI-SNR} & \textbf{Si-SNRi} & \textbf{\textit{OVRL}} & \textbf{\textit{SIG}} & \textbf{\textit{BAK}} & \thead{\textbf{Power proxy}\\ {[M-Ops/s]}} & \thead{\textbf{PDP proxy}\\ {[M-Ops]}} & \textbf{Params} \\
    \hline
    SDNN baseline  \cite{timcheck2023intel}   & 32 & 12.5  & 4.88 & 2.71 & 3.21 & 3.46 & 14.52 & 0.46 & 0.465M \\
    Spiking-FullSubNet (Large) \cite{hao2024audio}                    & 32 & 14.80 & 7.43 & 3.03 & 3.33 & 3.96 & 74.10 & 2.37 & 1.29M \\
    Spiking S4  \cite{du2023spiking}          & 32 & 14.58 &  7.21  & 2.85 & 3.21 & 3.74 &  --   &  --  & 0.53M \\
    Spiking-FullSubNet (Small) \cite{Intel2023DNSPage}                     & 32 & 13.89 & 6.52 & 2.97 & 3.28 & 3.93 & 29.24 & 0.94 & 0.953M \\
    CTDNN \cite{Intel2023DNSPage}                          & 32 & 13.52 & 6.59 & 2.97 & 3.32 & 3.86 & 61.37 & 0.49 & 0.90M\\
    \hline
    DPSNN (4.0-second inputs)                       & 5  & 14.54  & 7.18 & 2.88 & 3.27 & 3.72  & 175.2 (208.0)  & 0.88 (1.04) & 1.32M (1.40M) \\
    DPSNN (5.0-second inputs)                       & 5  & 14.60  & 7.23 & 2.89 & 3.29 & 3.71  & 180.4 (213.1)  & 0.90 (1.07) & 1.32M (1.40M) \\
    \hline
    \multicolumn{5}{l}{}
    \end{tabular}
\label{tab:dns}
\end{center}
\end{table*}



\section{Conclusion}
Speech enhancement is crucial for improving spoken communication in noisy environments, with applications ranging from automatic speech recognition to telecommunications. Recognizing the challenges posed by power constraints and the need for low latency in these domains, we explored the potential of neuromorphic algorithms through SNNs. 

Our research addresses the current limitations of effective SNN solutions, which typically impose substantial latency due to long sampling windows. Drawing inspiration from the efficacy of low-latency DPRNNs in deep learning, we have developed the DPSNN, a novel two-phase time-domain SNN framework. In this framework, the first phase uses SCNNs to capture global contextual information, while the second phase employs SRNNs to focus on frequency-related features. In addition, the threshold-based activation suppression combined with $L_1$ regularization loss are applied to specific non-spiking layers to further enhance energy efficiency of DPSNNs.

Evaluating on benchmark datasets such as VCTK and Intel DNS, our approach achieves significantly lower latency, approximately 5ms, compared to current solutions, while maintaining excellent SNR, perceptual quality, and energy efficiency. The model thus represents a significant advancement in speech enhancement techniques, promising improved communication experiences and efficiency across a wide range of applications.

\section{Acknowledgments}
This work was inspired by the discussions in the The CapoCaccia Workshops toward Neuromorphic Intelligence 2023. The authors also express their appreciation to Dr. Bojian Yin for his insightful recommendations concerning SNN training, to Dr. Guangzhi Tang and Dr. Paul Detterer for their instructive suggestions on model sparsity, and to Haohui Zhang for his valuable help regarding the figures. TS is supported by NWO-NWA grant NWA.1292.19.298. SB is supported by the European Union (grant agreement 7202070 ``HBP'').

\bibliographystyle{unsrt}
\bibliography{main}

\end{document}